\documentclass[
 preprint,
 amsmath,amssymb,
 aps,
 prl
superscript,
superscriptaddress,
]{revtex4-1}
\usepackage{graphicx} 
\usepackage{graphicx}
\usepackage{dcolumn}
\usepackage{bm}
\usepackage{here}
\usepackage{braket}
\usepackage{url}
\usepackage{comment}
\usepackage{ulem}
\usepackage{multirow}
\usepackage{color}

\newcommand{\tc}[1]{#1}

\usepackage[pdftex, pdftitle={Article}, pdfauthor={Author},
colorlinks=true,
linkcolor=blue,
citecolor=blue,
urlcolor=blue]{hyperref}

\usepackage{xcolor}
\colorlet{red}{black}

\bibpunct[, ]{[}{]}{,}{n}{,}{,}

\makeatother

\begin{document}

\title{Unconventional magnetic field response of X-ray magnetic circular dichroism in $\alpha$-Fe$_2$O$_3$}

\author{Yuta Ishii}
\email[Correspondence email address: ]{ISHII.Yuta@nims.go.jp}
\affiliation{Center for Basic Research on Materials (CBRM), 
National Institute for Materials Science (NIMS), Tsukuba 305-0047, Japan}

\author{Norimasa Sasabe}
\affiliation{Center for Basic Research on Materials (CBRM), 
National Institute for Materials Science (NIMS), Tsukuba 305-0047, Japan}

\author{Yuichi Yamasaki}
\email[Correspondence email address: ]{YAMASAKI.Yuichi@nims.go.jp}
\affiliation{Center for Basic Research on Materials (CBRM), National Institute for Materials Science (NIMS), Tsukuba 305-0047, Japan}
\affiliation{RIKEN Center for Emergent Matter Science (CEMS), Wako 351-0198, Japan}
\affiliation{International Center for Synchrotron Radiation Innovation Smart, Tohoku University, Sendai 980-8577, Japan}

\begin{abstract}
Altermagnets are compensated collinear magnets that break time-reversal symmetry without net magnetization, enabling unconventional magneto-optical responses. 
Here, we experimentally demonstrate X-ray magnetic circular dichroism (XMCD) in hematite $\alpha$-Fe$_2$O$_3$ 
for X-rays incident along the [111] direction, perpendicular to the applied magnetic field and the weak ferromagnetic moment induced by the Dzyaloshinskii--Moriya interaction.
The XMCD amplitude reaches approximately 10 \% of the polarization-averaged X-ray absorption intensity, 
whereas it is nearly suppressed for $\mathbf{H}\parallel[1\bar{1}0]$ of 200 mT.
This unique behavior is consistent with field-driven changes between configurations associated with different magnetic symmetries.
Our results establish that XMCD in the altermagnet $\alpha$-Fe$_2$O$_3$ exhibits a fundamentally distinct response compared to that of weak ferromagnetism.
This study addresses how XMCD in altermagnets responds to external magnetic fields compared to conventional ferromagnets.
\end{abstract}

\maketitle

Altermagnetism, a collinear compensated magnetic phase with anisotropic spin symmetry, 
has recently emerged as a new frontier in spintronics and condensed matter physics 
\cite{Smejkal2020_SciAdv,Smejkal2022_PRX}. 
Distinguished from conventional N\'{e}el antiferromagnetism (AFM) and ferromagnetism (FM), 
altermagnets break time-reversal symmetry without net magnetization, 
driving spin splitting in momentum space with $d$-, $g$-, and higher-order wave symmetry 
\cite{Noda2016_PhysChemChemPhys,Okugawa_AM2018,Ahn2019PRB,Naka2019NatCom,Smejkal2020_SciAdv}. 
These materials exhibit remarkable transport and optical responses, 
including anomalous Hall and Nernst effects 
\cite{Smejkal2020_SciAdv,Gonzalez2023,Reichlova2024,Sato2024,Sheoran2025}, 
as well as unconventional magneto-optical dichroism 
\cite{Hariki2024,Amin2024,Peter2025,Galindez2025}.
The essential ingredient is the crystal symmetry that connects opposite spin sublattices by proper rotations or mirror operations rather than simple translations or inversions, 
leading to momentum-dependent spin splitting even in the absence of net magnetization. 
The symmetry of the spin splitting directly reflects the underlying crystalline point group and determines the anisotropy of spin-dependent responses.

Recently, it has been theoretically and experimentally demonstrated that X-ray magnetic circular dichroism (XMCD) $\Delta \mu$, 
defined as the difference in the X-ray absorption spectrum (XAS) between right- and left-circularly polarized light, 
can acquire a finite signal even in AFMs,
including several altermagnets
\cite{Yamasaki2020, XMCD_Kimata2021,Sasabe_PRL_Mn3Sn2021,Hariki2024,Amin2024,Galindez2025}. 
Although XMCD is traditionally regarded as a probe of net magnetization, 
it is symmetry-allowed in AFMs 
when the electronic structure is characterized by the symmetry of a cluster magnetic dipole.
X-ray absorption is an intra-atomic excitation process, 
and therefore, even in an AFM, it probes the magnetic dipole components that break time-reversal symmetry.
It is suggested that 
this magnetic dipole can be represented 
by a ferroic-component anisotropic magnetic dipole (AMD) moment \cite{Yamasaki2020,sasabe2023ferroic,hayami2024unified, YamasakiSTAM2025}. 
\tc{The AMD moment arises from the coupling between the spin and anisotropic charge distributions and is expressed as $\bm{t} = \sum_i \bm{Q}^i \bm{s}^i$, where $\bm{s}$ and $\bm{Q}$ denote the spin and electric quadrupole tensor, respectively \cite{Oguchi2004}. 
Thus, the AMD moments can align ferroically even in an antiferromagnetic spin configuration 
if the antiferromagnetic sublattice sites are related not by a simple translational operation but by mirror symmetry or rotational operations, and the
electric quadrupole moments at the respective sites have opposite signs.}
The total AMD corresponds to the physical quantity of the magnetic dipole term $T_z$ operator that appears in the XMCD sum rules and has long been recognized in that context \cite{Thole1992, Carra1993}. 
Consequently, XMCD is an effective tool 
for investigating 
the breaking of time-reversal symmetry in altermagnets.
Despite several XMCD investigations on altermagnets, understanding the response of XMCD to an external magnetic field remains crucial for exploring the functionalities of altermagnets.

Among the candidate altermagnetic materials, hematite ($\alpha\text{-Fe}_2\text{O}_3$) stands out as a prototypical $g$-wave altermagnet. 
At room temperature, it crystallizes in the space group of trigonal $R\bar{3}c$ and exhibits a collinear AFM order (shown in Fig.~1).
The crystallographic point group is $\bar{3}m$ (Schoenflies notation $D_{3d}$), characterized by a threefold rotation axis [111]$_\text{rh}$ and a glide plane parallel to it ($c$-glide).
Note that, throughout this paper, crystallographic directions are expressed in the rhombohedral setting.
It is a well-known AFM with a high N\'eel temperature ($T_N \sim 950$ K) and undergoes a Morin transition at $T_M \approx 250$ K; 
in its high-temperature phase ($T > T_M$), it 
exhibits a weak ferromagnetic (wFM) moment 
within the (111) plane induced by the Dzyaloshinskii-Moriya (DM) interaction \cite{Dzyaloshinskii1958,Moriya1960}.
In addition, 
an even weaker magnetization 
along the [111] direction 
has been reported
\cite{Flanders1972}.
At room temperature, the N\'eel vector $\mathbf{N}$ lies within the $(111)$ plane.
To describe the in-plane orientation of the N\'eel vector,
we introduce an orthonormal coordinate system with
$\hat{\boldsymbol{x}}\parallel[1\bar{1}0]$,
$\hat{\boldsymbol{y}}\parallel[11\bar{2}]$, and
$\hat{\boldsymbol{z}}\parallel[111]$, 
using rhombohedral indices throughout. 
The unit N\'eel vector is then written as $\boldsymbol{n}
= \frac{\boldsymbol{N}}{|\boldsymbol{N}|}
= (\cos\phi_N,\sin\phi_N,0)$,
where $\phi_N$ is measured in the $(111)$ plane from
$\hat{\boldsymbol{x}}$ toward $\hat{\boldsymbol{y}}$.
The magnetic symmetry depends on the orientation of $\mathbf{N}$ relative to the crystallographic mirror planes.
As shown in Figs. 1(c) and 1(d), when $\mathbf{N} \parallel \hat{\boldsymbol{x}}$, the magnetic point group (MPG) is $2'/m'$, whereas
 for $\mathbf{N} \parallel \hat{\boldsymbol{y}}$, it becomes $2/m$.  
The distinction arises from whether the relevant mirror operation is preserved as a pure spatial symmetry ($m$) or is combined with time reversal ($m'$).
Although both MPGs break time-reversal symmetry, $2'/m'$ allows symmetry breaking along both the in-plane wFM moment direction $\langle 11\bar{2} \rangle$
and the $\langle111\rangle$ directions, whereas $2/m$ confines it to the in-plane $\langle 1\bar{1}0 \rangle$ direction, excluding any $\langle111\rangle$ component.
These distinct magnetic symmetries are characterized by the alignment of the AMD moments. 
In the case of $\textbf{N}||x$, 
the AMD moments exhibit ferroic alignment, 
whereas in $\textbf{N}||y$, they exhibit antiferroic alignment
(see Supplementary Material).
In contrast, below $T_M$, when $\mathbf{N}\parallel z$, 
MPG becomes $\bar{3}m$, under which magnetic dipole moments are forbidden.

From the viewpoint of magnetic symmetry, 
an out-of-plane XMCD response is allowed for the $2'/m'$ reference configuration of $\alpha$-Fe$_2$O$_3$. 
Its local trigonal crystal field and collinear AFM structure closely resemble those of $\alpha$-MnTe \cite{Hariki2025, Amin2024, sasabe2025}. 
Experimentally, photoemission electron microscopy (PEEM) XMCD has recently been reported in Ti-doped $\alpha$-Fe$_2$O$_3$ thin films \cite{Galindez2025}.
In addition, symmetry-driven magneto-optical Kerr effects (MOKE) have also been proposed and experimentally explored \cite{Luo_2026,Pan2026}. 
These studies establish that optical responses are symmetry-allowed and potentially observable in $\alpha$-Fe$_2$O$_3$. 
However, how the magnitude and sign of XMCD in hematite depend on the N\'eel vector and on the strength and in-plane direction of an applied magnetic field has not been established experimentally.
In this study, we systematically measure the dependence of XMCD in $\alpha$-Fe$_2$O$_3$ on the strength and in-plane direction of an applied magnetic field. 
We combine these measurements with linear dichroism, symmetry analysis, and multiplet calculations to investigate the associated changes in antiferromagnetic order.
Our results demonstrate a symmetry-sensitive optical response, providing a foundation for the detection and manipulation of antiferromagnetic states in spintronic materials.

XMCD measurements around the Fe $L_{2,3}$ edges ($2p\rightarrow 3d$) 
were performed at undulator beamline BL-16A of the Photon Factory (KEK, Japan), using an XMCD end-station equipped with a 5~T superconducting magnet.
The sample was a natural $\alpha$-Fe$_2$O$_3$ single-crystal substrate purchased from SurfaceNet GmbH (Germany). 
It measured $5 \times 5 \times 0.5~\mathrm{mm}^3$ and had a (111) surface, corresponding to the (0001) plane in the hexagonal setting.
The in-plane $[1\bar{1}0]$ direction was marked on the substrate by the supplier.
One side was mirror-polished, and the surface orientation accuracy was $\pm 0.5^\circ$.
Before the measurements, the mirror-polished surface was cleaned with ethanol.
The spectra were recorded in total electron yield (TEY) mode. 
Electrical contact between the sample surface and the copper sample holder was established using conductive carbon tape for the TEY measurements.
The nominal energy resolution of BL-16A is \(\Delta E/E \sim 2\times10^{-4}\), corresponding to approximately 0.14 eV at the Fe \(L_{2,3}\) edges.
The external magnetic field ($\mathbf{H}_\text{ext}$) is applied parallel to the X-ray incident direction $\textbf{k}$.
The experimental setup is equipped with a motorized single-axis rotation stage. 
The sample was mounted such that the [$11\bar{2}$] direction is parallel to the rotation axis. 
At the rotation angle $\theta = 0^\circ$, $\mathbf{H}_\text{ext}$ is applied along the [111] direction, 
whereas at $\theta = 90^\circ$, the field is approximately directed toward the [$1\bar{1}0$] axis.
To enhance measurement sensitivity and suppress systematic drifts, 
the helicity of the circularly polarized X-rays was switched between left and right polarization at a frequency of 10 Hz \cite{Amemiya2013BL16A}.
The degree of circular polarization during helicity switching was approximately 90 \% \cite{Yasuda2023Mn4N}.
The XMCD signal 
was detected using a lock-in amplifier synchronized to the polarization switching frequency, enabling high-precision extraction of the dichroic signal.
Most measurements were performed at room temperature. 
To examine whether the XMCD signal vanishes below the Morin transition temperature, additional measurements were performed at 50~K using a helium-flow cryostat [see Supplementary Material].
The measurements with a magnetic field applied perpendicular to the incident X-ray beam were performed using the four-pole magnet XMCD equipment installed at NanoTerasu BL14U \cite{Ishii2025}.

Figure~2(a) shows the XAS and XMCD spectra around the Fe $L_{2,3}$ edges of $\alpha$-Fe$_2$O$_3$(111).
The upper panel presents the polarization-averaged XAS spectrum $(I_{+}+I_{-})/2$, 
reflecting the unoccupied Fe $3d$ density of states under crystal-field and multiplet effects.
Here, $I_{\pm}$ indicates XAS intensity measured with right or left circular polarization.
Characteristic multiplet structures corresponding to the high-spin Fe$^{3+}$ ($3d^5$) configuration are clearly observed at the $L_3$ ($\sim 709$ eV) and $L_2$ ($\sim 722$ eV) edges.
The lower panel shows the XMCD spectrum defined as $I_{+}-I_{-}$, exhibiting the characteristic oscillatory structure typical of altermagnets, which is distinctly different from the conventional XMCD line shape observed in ferromagnets \cite{Yamasaki2025High-throughput}. 
In addition, 
the magnitude of this signal 
reaches approximately 10 \% of the XAS intensity, which is larger than that reported for other altermagnetic systems \cite{Hariki2024,Amin2024}.
These results indicate that the observed XMCD signal along the [111] direction is not primarily governed by the previously reported weak magnetization component \cite{Flanders1972}, but instead originates predominantly from the anisotropic electronic structure \tc{reflecting the altermagnetic symmetry}.
For each remanent spectrum in Fig.~2(a), the sample was first oriented so that the applied field was parallel to the in-plane $[1\bar{1}0]$ direction. 
A preparation field $H_{\mathrm{pre}}=+5$ or $-5$~T was applied, after which the field was reduced to zero. 
The sample was then rotated to the normal-incidence geometry, $\mathbf{k}\parallel[111]$, and the XAS and XMCD spectra were acquired at zero field.
This preparation and acquisition sequence was repeated for the opposite sign of $H_{\mathrm{pre}}$.
The reversal of the remanent XMCD signal was reproduced upon repeating the preparation sequence with opposite field polarities.
Reversing the sign of the preparatory magnetic field reverses the zero-field XMCD signal, demonstrating that the AFM domains are controllably selected by the pre-applied field.
\tc{It should be noted that this XMCD signal vanishes below the Morin transition temperature because the magnetic dipole is forbidden by the $\bar{3}m$ MPG symmetry (see Supplementary Material}).

Since the DM-induced wFM moment lies within the $(111)$ plane and is perpendicular to $\mathbf{N}$, 
an in-plane magnetic field $\mathbf{H}_{\text{pre}}||[1\bar{1}0]$ is expected to preferentially select domains with $2/m$ symmetry 
($\mathbf{N}\parallel[11\bar{2}]$), as shown in Fig.~1(d). 
However, in this symmetry, the XMCD 
along the [111] direction should vanish. 
In contrast, for the $2'/m'$ configuration, the 
\tc{AMD} can possess a finite projection along the $[111]$ direction, making XMCD symmetry-allowed.
Therefore, the finite remanent XMCD is consistent with a configuration close to the $2'/m'$ reference state, but does not uniquely determine the magnetic point group.
Under applied $\mathbf{H}_{\text{pre}}\parallel[1\bar{1}0]$, the N\'eel vector belonging to the $2'/m'$ MPG can, in principle, take either $\mathbf{N}\parallel[01\bar{1}]$ or $\mathbf{N}\parallel[10\bar{1}]$, 
since these two configurations form equal angles with the applied field and are
energetically degenerate in the ideal geometry. 
However, in the actual experiment, a slight misalignment of the sample inevitably introduces a small deviation of $\mathbf{H}_{\text{pre}}$ from the exact $[1\bar{1}0]$ direction. 
Such a minute symmetry-breaking field component lifts the degeneracy between the two domain states and preferentially selects one of the two possible $\mathbf{N}$ orientations.
The XMCD signal retains its sign even after the magnetic field is removed. 
This persistence of the dichroic response at 0 T indicates that the selected AFM domain remains stable in zero field, 
demonstrating that $\mathbf{N}$ orientation is isothermally determined by $\mathbf{H}_{\text{pre}}$ and preserved as a remanent AFM domain state.

The spectral line shape and experimental configuration of the observed XMCD ($\mathbf{k}\perp\mathbf{M}_\text{wFM}$) differ markedly from that expected from a net wFM moment, indicating a nontrivial origin rooted in crystal symmetry, especially the local distortion of the ligand field.
To clarify this point, we calculated the XAS spectra using the \textsc{EDRIXS} code \cite{WANG2019}, based on the electronic states obtained by exact diagonalization of the Hamiltonian, incorporating a crystal field that reflects the actual ligand positions obtained from the Materials Project database (mp-19770, primitive cell \cite{Jain2013,Horton2025}), as well as spin--orbit coupling (SOC) and intra-atomic
Coulomb interactions.
The Slater integrals were reduced to 70 \% of their atomic values, and the calculated spectra were broadened to account for the core-hole lifetime. 
The numerical parameters for the initial and core-excited states are given in Supplementary Material.
Figure~2(b) shows the calculated XAS and XMCD spectra at the Fe-$L_{2,3}$ edges. 
The overall undulating spectral shape at both the $L_3$ and $L_2$ edges is well reproduced by the calculation.
Since AMD contributions have alternating signs
\cite{sasabe2025}, 
the resulting XMCD exhibits the pronounced oscillatory structure that is characteristic of
XMCD of altermagnets.
Beyond reproducing the spectral line shape, the calculation quantitatively accounts for the XMCD intensity relative to the XAS.
The remaining discrepancy near the higher-energy shoulder of the main peak may partly arise from energy-dependent broadening within the multiplet manifold, Fe off-centering and associated hybridization with odd-parity orbitals such as Fe $4p$, or Fe--O charge-transfer configurations, such as $3d^6\underline{L}$ ($\underline{L}$ denotes a ligand hole), which are not included in the present calculation.

In contrast, the XMCD spectrum for $\mathbf{k}||[\bar{2}11]$ calculated from the wFM component is shown in Fig. 2(b). 
In the calculation, the canting angle $0.0554^\circ$ \cite{Hill2008ChemMater} is used to estimate the magnitude of $\mathbf{M}_\text{wFM}$, 
and the corresponding XMCD response with $\mathbf{k}||\mathbf{M}_\text{wFM}$ is evaluated within the same multiplet framework.
In particular, the calculated wFM spectrum does not reproduce the characteristic peak features of the experimental spectrum, 
and the overall magnitude of the calculated signal is more than one order of magnitude smaller than the observed XMCD intensity.
These results demonstrate that the experimentally observed XMCD cannot be attributed to the DM-induced wFM moment.

Figure 3(a) shows the $\mathbf{H}_{\text{ext}}$ dependence of XMCD spectra taken with the sample tilted such that $\mathbf{H}_{\text{ext}}$ and $\mathbf{k}$ are oriented $\theta= 30^\circ$ away from the surface normal.
Here, the tilt angle $\theta$ is defined between the surface normal and the incident X-ray beam, which is parallel to $\mathbf{H}_{\mathrm{ext}}$.
By rotating the sample around the $y||[11\bar{2}]$ axis, the magnetic field acquires an in-plane component along the $x||[1\bar{1}0]$ direction, as shown in the inset of Fig.~3(a).
Prior to the measurement, a preparation field of $H_{\mathrm{pre}}=+5$~T was applied along the $\hat{x}||[1\bar{1}0]$ direction to select the magnetic domains.
The field was then reduced to zero, and an XMCD spectrum was acquired at normal incidence ($\theta=0^\circ$).
The sample was subsequently rotated to $\theta=30^\circ$, where spectra were acquired first at zero field and then at $H_{\mathrm{ext}}=5$~T.
For the geometry of Fig.~3(a), the observed XMCD is expressed as 
$\Delta\mu_{\mathrm{obs}}(\theta)
= \Delta\mu_x\sin\theta
+ \Delta\mu_z\cos\theta$.
At zero applied field, tilting the sample from $\theta=0^\circ$ to $30^\circ$ reduces the XMCD amplitude without appreciably changing the spectral line shape.
The reduction is consistent with the geometrical projection of the surface-normal response, 
$\Delta\mu_{\mathrm{obs}}(30^\circ)\simeq \Delta\mu_z\cos30^\circ$.
Thus, no distinct in-plane XMCD contribution $\Delta\mu_x$ is resolved in the prepared remanent state within the sensitivity of the present measurement.
Upon applying the magnetic field at $\theta=30^\circ$, 
the XMCD amplitude decreases while its spectral line shape remains largely unchanged. 
This suppression is inconsistent with a simple enhancement of $\Delta \mu_x$ component from the wFM contribution.
Figures 3(b) and 3(c) display the magnetic-field dependence of the XMCD intensity at the $L_2$ edge ($E = 722.3$ eV) for $\theta = 0^\circ$ and $\theta = 30^\circ$, respectively. For $\theta = 0^\circ$, the XMCD exhibits a clear hysteresis loop, reaching saturation at a magnetic field of approximately 200 mT while retaining a finite remanent signal. In contrast, the magnetic-field dependence for $\theta = 30^\circ$ is markedly different from that observed for $\theta = 0^\circ$. The XMCD signal exhibits a sharp sign reversal across zero field, followed by a gradual suppression.
The dynamic response is further demonstrated in Fig.~3(d), 
where the magnetic field is linearly swept over time. 
The XMCD signal follows the field variation reproducibly and switches sign whenever the field polarity changes, 
confirming that the observed dichroism is directly coupled to the field-controlled magnetic domain state.

This unique behavior under an in-plane magnetic field is inconsistent with a conventional ferromagnetic hysteresis loop and instead suggests that the XMCD originates from a field-selectable antiferromagnetic order parameter.
In the tilted geometry, the field has both an in-plane component
$H_{\parallel}=H_{\mathrm{ext}}\sin\theta$ along
$\hat{x}\parallel[1\bar{1}0]$ and a surface-normal component
$H_{\perp}=H_{\mathrm{ext}}\cos\theta$ along
$\hat{z}\parallel[111]$.
As schematically illustrated in Fig.~3(e), coupling to the
weak-ferromagnetic moment and the antiferromagnetic spin
configuration may favor N\'eel-vector orientations closer to
$\hat{y}\parallel[11\bar{2}]$ as $H_{\parallel}$ increases,
in which the AMD moments exhibit antiferroic alignment (see Supplementary Material).
Such a reorientation would reduce $\Delta \mu_z$ and is consistent with the observed suppression.
A change in the populations of antiferromagnetic domains may
also contribute.
Such a magnetic field control of spin reorientation
has also been reported in a previous XMLD study \cite{Suturin2021}.
When the field is reduced to zero, the XMCD signal recovers, demonstrating recovery of the remanent dichroic response.
For $\theta=0^\circ$, the applied field has no in-plane component in the ideal geometry, and the XMCD instead exhibits a hysteretic response. 
Unlike the pronounced suppression observed at $\theta=30^\circ$, a substantial XMCD signal persists up to the highest field measured,
although its sign changes with the field history.
This sign reversal may involve coupling to the very weak surface-normal magnetic moment reported for hematite \cite{Flanders1972}, an in-plane field component arising from slight sample misalignment, or both.

The effect of the in-plane magnetic field is further investigated using an XMCD setup capable of applying an in-plane magnetic field \cite{Ishii2025}.
Figures 4(a) and 4(b) show the XMCD measured along the [111] direction around the $L_2$ edge under magnetic fields applied along the [1$\bar{1}$0] direction. 
For the measurement, an in-plane magnetic field of +200~mT was first applied along $[1\bar{1}0]$.
The field was then reduced to +2.5~mT, where the Fe $L_2$-edge XMCD spectrum was acquired for Fig. 4(a). 
Subsequently, the field was reversed to $-2.5$~mT and a second spectrum was acquired.
For Fig.~4(b), the XMCD intensity was recorded at a fixed photon energy of 722.3~eV while the in-plane magnetic field was swept from +200 to $-200$~mT and then back to +200~mT.
Throughout these measurements, the incident X-ray beam was parallel to [111].
The XMCD signal reverses its sign even under a weak magnetic field of 2.5 mT, but is nearly suppressed
at an applied field of approximately 100~mT.
This result suggests that the sign of the N\'eel vector can be reversed by a weak in-plane magnetic field, which can be clearly observed through the AMD moment in XMCD. 
In addition, this behavior is consistent with a field-induced change from the dichroic magnetic state to the $\textbf{N}||\hat{y}$ state at an in-plane field of order 100~mT.
This field is lower than the in-plane component of the tilted field associated with XMCD suppression as shown in Fig.~3(c).
One possible explanation is that the surface-normal field component $H_\perp$ in the tilted geometry favors the dichroic magnetic state $\textbf{N}||\langle1\bar{1}0\rangle$ and thus shifts the apparent suppression field.
Further measurements are needed to test this possibility and determine the magnetic switching pathway.

Figure 4(c) shows the XMCD intensity at $E = 722.3$ eV as a function of the in-plane magnetic-field direction.
Here, the magnitude of the in-plane magnetic field was fixed at 90~mT, 
and the sample was rotated about its [111] axis to vary the field direction relative to the crystal axes.
The incident X-ray beam remained parallel to [111] throughout the measurement.
The XMCD exhibits a threefold rotational symmetry depending on the magnetic field direction. 
To examine whether the change in XMCD is accompanied by a reorientation of the N\'eel vector, 
we measured X-ray magnetic linear dichroism (XMLD) spectra at $\phi=0^\circ$ and $\phi=90^\circ$ under a fixed in-plane magnetic field of 90~mT [Fig.~4(d)]. 
The dichroism was obtained from the difference between XAS spectra measured with linear polarizations along the $[1\bar{1}0]$ and $[11\bar{2}]$ directions. 
The spectra exhibit opposite signs for the two field orientations. This reversal of the XMLD signal provides evidence that the N\'eel vector reorients when the direction of the in-plane field is changed. 
Together with the angular dependence of XMCD in Fig.~4(c), it supports field-controlled selection of antiferromagnetic states.
This behavior is consistent with the three fold symmetry of the AMD moment 
with respect to the in-plane N\'eel vector orientation, $\Delta \mu \propto \cos(3\phi_N)$ \cite{sasabe2025}, 
because the AMD moment arises from the anisotropic electronic structure and therefore strongly reflects the threefold rotational symmetry of the trigonal crystal field.
A complementary interpretation is provided by the optical Hall-vector description of XMCD and the recent multipolar framework, 
in which a leading out-of-plane component with the same threefold angular dependence, $h_z \propto n_x(n_x^2-3n_y^2)=\cos(3\phi_N)$, is allowed \cite{Xie2026,KunesPRB25,Xiao2025AnomalousHall}.
We note that the $g$-wave classification of $\alpha$-Fe$_2$O$_3$ refers to the symmetry of the momentum-space spin splitting, whereas the threefold symmetry discussed here reflects the N\'eel-vector dependence of the XMCD response rather than directly probing the momentum-resolved spin splitting.

Such an unconventional magnetic-field response further provides compelling evidence that $\alpha$-Fe$_2$O$_3$ hosts a magnetic structure characterized by time-reversal symmetry breaking of altermagnets and distinct from that of conventional weak ferromagnetism.
In addition, the pronounced contrast between the field-selected states suggests that symmetry-sensitive X-ray dichroism can serve as a readout of antiferromagnetic domain states in hematite.
The ability to modify this contrast with small in-plane magnetic fields may be useful for exploring the control and detection of compensated magnetic order in spintronic
materials.

In conclusion, we have experimentally demonstrated the 
unconventional XMCD response 
in the altermagnet $\alpha$-Fe$_2$O$_3$ at room temperature.
The observed oscillatory line shape, 
characteristic of altermagnetic symmetry, 
is well reproduced by full-multiplet calculations based on exact diagonalization of the realistic electronic Hamiltonian.
Furthermore, the combined XMCD and XMLD results demonstrate field-controlled reorientation of the antiferromagnetic order and reversible changes in the XMCD response, 
consistent with switching between states characterized by the $2'/m'$ and $2/m$ magnetic point groups.
These findings establish a basis for magnetic-field control and symmetry-sensitive optical readout of altermagnetic states in room-temperature materials.
Extending field-dependent XMCD measurements to other altermagnets and comparing them with magneto-optical Kerr measurements will help assess the generality of this approach and its potential for spintronic devices.

The authors are grateful to K. Ozawa, K. Amemiya, and Y. Kotani for their support of the experiments 
and to T. Arima for insightful discussions.
The synchrotron radiation experiments were performed in the Photon Factory with the approval of the Photon Factory Program Advisory Committee (Proposal Nos. 2024G615 and 2024G630) and 
performed in the NanoTerasu with the approval of the Japan Synchrotron Radiation Research Institute (JASRI) (Proposal No. 2026A9408).
This work was supported by JSPS KAKENHI (Grants No. JP19K23590, JP19H04399, JP20K20107, JP23K17145, JP24K03205, JP24H01685, JP25K03387),  by PRESTO (JPMJPR2102), and CREST (JPMJCR1861 and JPMJCR2435), Japan Science and Technology Agency (JST).

\bibliographystyle{tfnlm}
\bibliography{reference}
\clearpage

\begin{figure}[tbp]
\begin{center}
\label{fig:Structure}
\includegraphics[width=8.5cm]{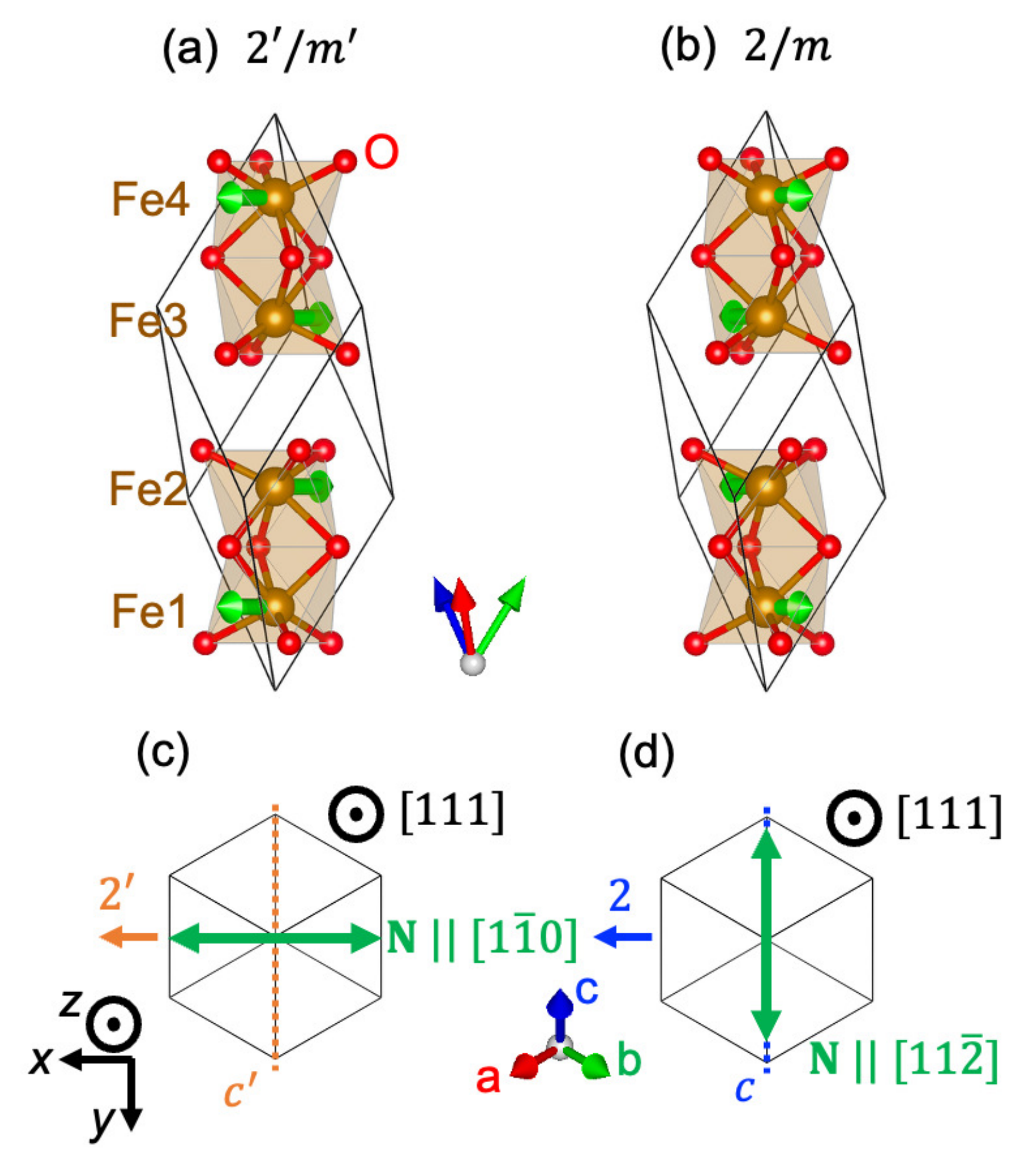}
\caption{
Crystal and magnetic structure of $\alpha$-Fe$_2$O$_3$ for 
(a) the N\'eel vector $\mathbf{N}\parallel[1\bar{1}0]$ and
(b) $\mathbf{N}\parallel[11\bar{2}]$. 
The corresponding magnetic point groups are 
$2^{\prime}/m^{\prime}$ and $2/m$, respectively.
The structures are visualized using VESTA \cite{Momma_2011}. 
Schematic view along the $[111]$ direction of $\mathbf{N}$, 
$c$- and $c^{\prime}$-glide planes
as well as the twofold rotation axes 
($2$ and $2^{\prime}$) 
for (c) $\mathbf{N}\parallel[1\bar{1}0]$
and (d) $\mathbf{N}\parallel[11\bar{2}]$, respectively.
} 
\end{center}
\end{figure}
\begin{figure}[tbp]
\begin{center}
\label{fig:spectra}
\includegraphics[width=8.5cm]{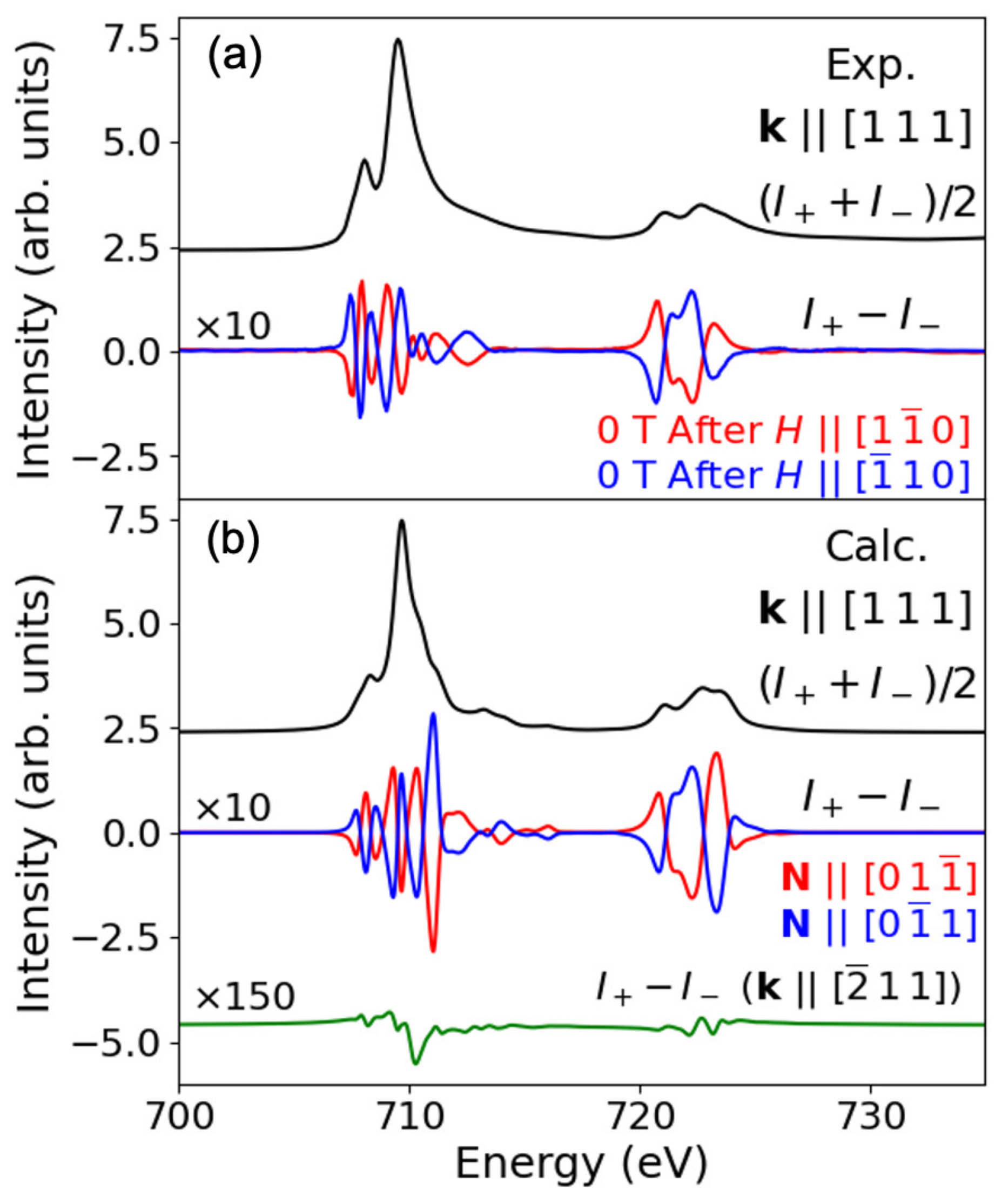}
\caption{
(a) X-ray absorption spectroscopy (XAS) and X-ray magnetic circular dichroism (XMCD) spectra at the Fe $L_{2,3}$ edges for $\alpha$-Fe$_2$O$_3$(111). 
 The upper panel shows the polarization-averaged XAS spectrum, $(I_{+} + I_{-})/2$, measured in total electron yield (TEY) mode. 
The lower panel displays the XMCD signal, defined as $I_{+} - I_{-}$, obtained at 0 T after applying external magnetic fields of $\pm 5$~T along the [$1\bar{1}0$] direction. 
The red and blue curves correspond to measurements after applying $+5$~T and $-5$~T, respectively. 
The XMCD intensity is magnified by a factor of 10 for clarity.
(b) Calculated XAS and XMCD spectra.
Red and blue lines correspond to XMCD 
of $\mathbf{N}~||~[01{\bar1}]$ and 
$\mathbf{N}~||~[0{\bar1}1]$.
Green line presents XMCD spectrum calculation 
originating from $\mathbf{M}_{\rm wFM}$ along the [$\bar{2}11$]
direction.
The intensity is magnified by a factor of 150.} 
\end{center}
\end{figure}

\begin{figure*}[t]
\begin{center}
\label{fig:spectra_magnetic-field}
\includegraphics[width=17cm]{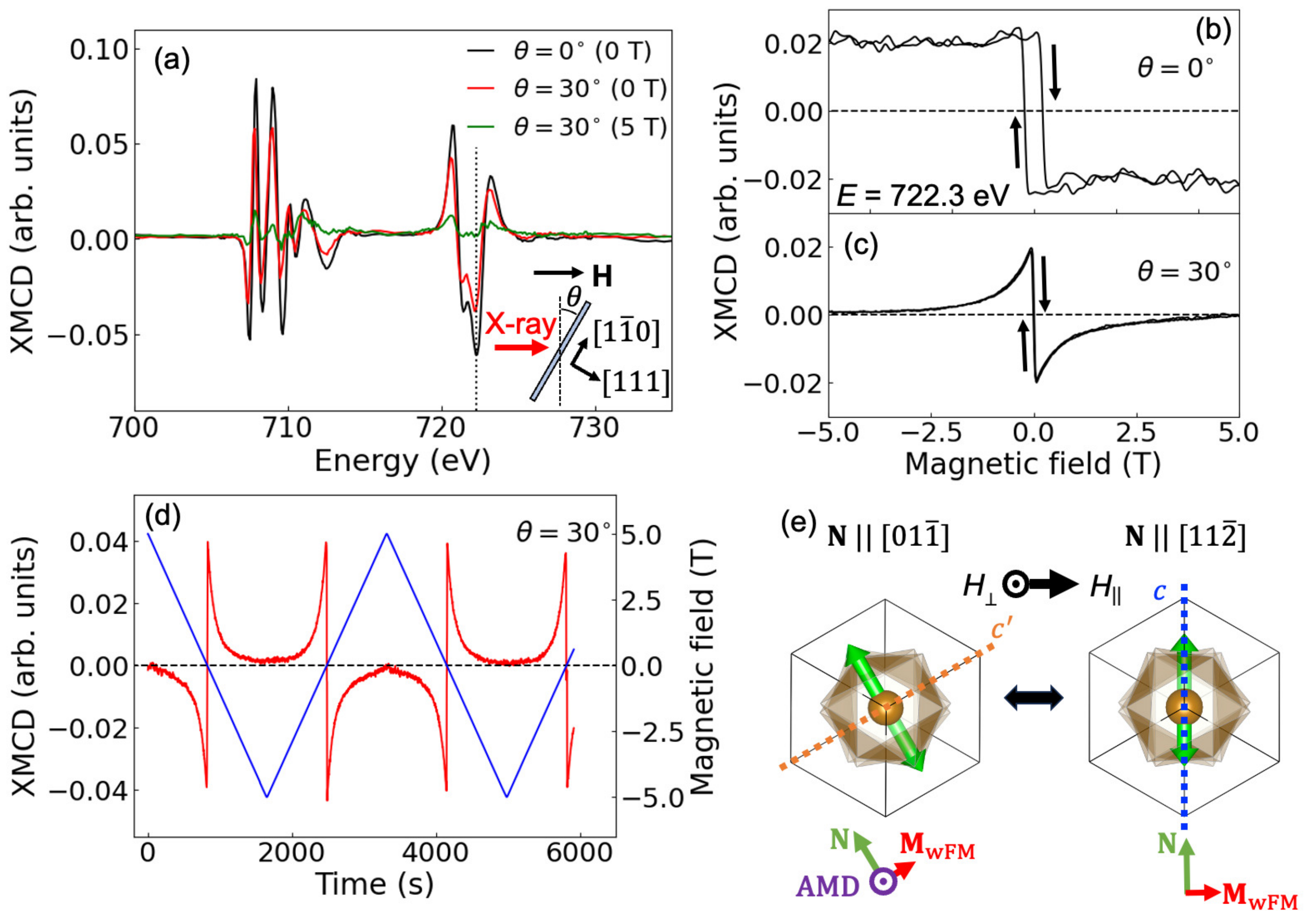}
\caption{(a) XMCD spectra measured around Fe $L_{2,3}$-edges under external magnetic field.
The tilt angle $\theta$ is defined between the surface normal and the incident X-ray beam, 
which is parallel to magnetic field.
A dashed vertical line marks 
$E = 722.3$ eV.
The inset illustrates experimental geometry.
Magnetic-field dependence of the XMCD intensity at $L_2$ edge ($E = 722.3$ eV) 
with (b) $\theta = 0^{\circ}$ and (c) $\theta = 30^{\circ}$. 
(d) Time evolution of the XMCD signal at 722.3 eV under  
a linearly varying magnetic field.
(e) Schematic illustration of magnetic structure variation 
between $\mathbf{N}\parallel[01\bar{1}]$
and $\mathbf{N}\parallel[11\bar{2}]$
in the presence of an in-plane $H_{\parallel}||[1\bar{1}0]$ and out-of-plane $H_{\perp}||[111]$ magnetic fields.
}
\end{center}
\end{figure*}

\begin{figure}[tbp]
\begin{center}
\label{fig:rotH}
\includegraphics[width=15cm]{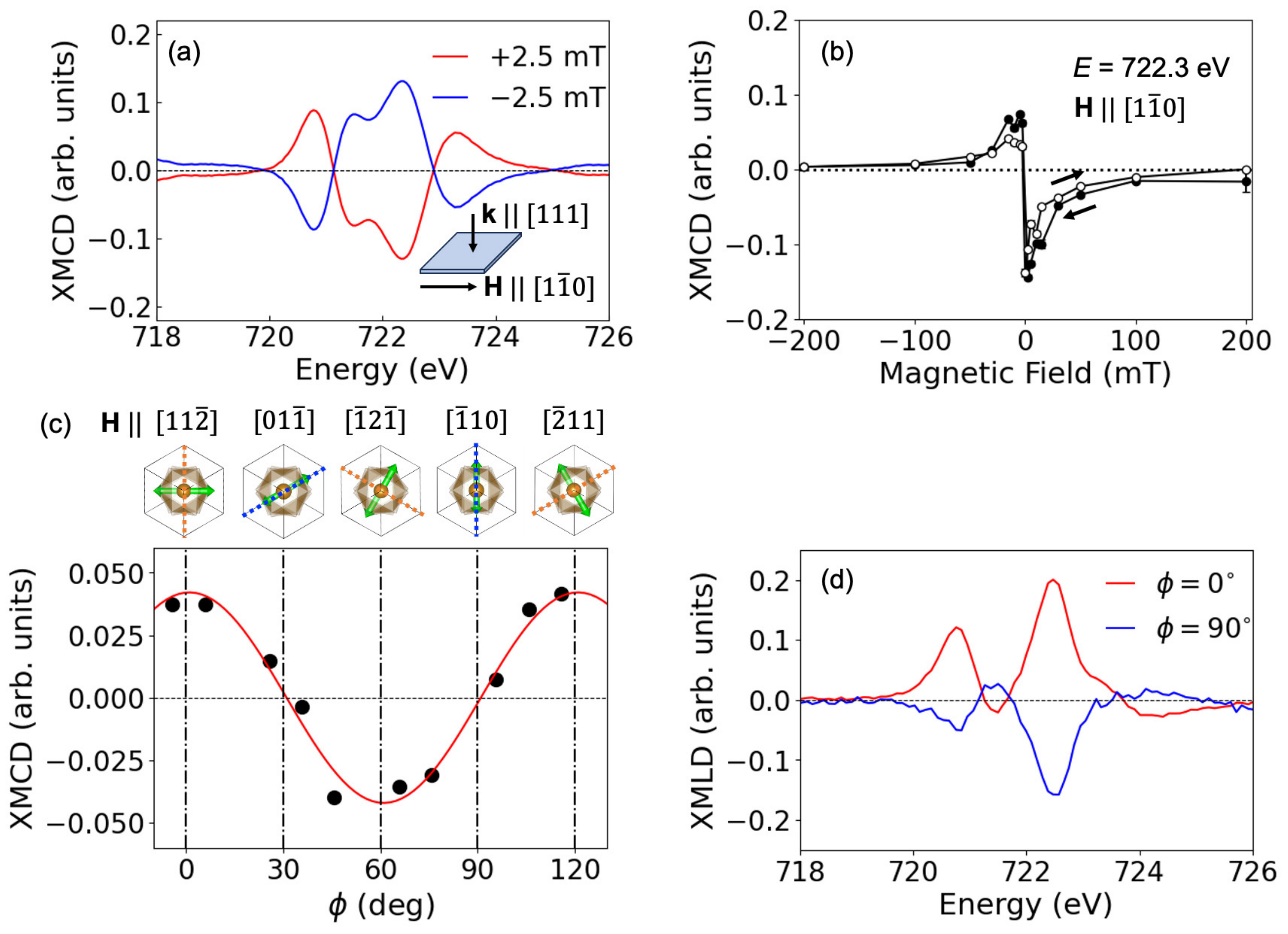}
\caption{(a) XMCD spectra around the $L_2$ edge measured along the [111] direction under magnetic fields of $H = \pm 2.5$ mT applied along the [$1\bar{1}0$] direction.
(b) Magnetic field dependence of XMCD intensity at $E = 722.3$ eV.
Filled and open circles represent the data obtained while decreasing and increasing the magnetic field, respectively.
(c) Variation of XMCD intensity at $E = 722.3$ eV as a function of the in-plane magnetic-field direction. $\phi$ denotes the angle between the magnetic-field direction and the [$11\bar{2}$] direction. 
(d) X-ray magnetic linear dichroism (XMLD) spectra around the \(L_2\) edge at \(\phi = 0^{\circ}\) and \(\phi = 90^{\circ}\). The XMLD is defined as the difference between the XAS spectra for incident X-rays linearly polarized along the [$1\bar{1}$0] and [$11\bar{2}$] directions.
The in-plane magnetic field was fixed at 90 mT for both (c) and (d).
}
\end{center}
\end{figure}

\clearpage
\newpage
\begin{center}
  {\Large\bfseries Supplemental Material
  for ``Unconventional magnetic field response of X-ray magnetic circular dichroism in $\alpha$-Fe$_2$O$_3$''}\\[0.5em]
\end{center}

\section{Electronic states and anisotropic magnetic dipole moments in $\alpha$-Fe$_2$O$_3$}

Here, the electronic states and 
anisotropic magnetic dipole (AMD) moments 
in $\alpha$-Fe$_2$O$_3$
are investigated.
The five degenerate $d$-orbitals of the Fe ion split into $e_g$ and $t_{2g}$ states due to the octahedral crystal field ($O_h$). 
A trigonal distortion reduces the symmetry to $D_{3d}$ and causes these degenerated states to split further: the $t_{2g}$ level splits into $a_{1g}$ and $e_g^\pi$ components.
Using the spherical harmonic basis $|m\rangle$, 
the $a_{1g}$ is represented by $\lvert 0 \rangle$, and the $e_g^\pi$ states are 
$|\pi_\pm \rangle = \cos\zeta|\pm 2\rangle \mp \eta_j \sin \zeta|\mp 1\rangle$
with $\zeta$ depending on the trigonal distortion.
The four inequivalent Fe sites are distinguished by the parameter $\eta_j$ at the $j$-th site, taking the values $\eta_1 = \eta _4 = 1$ and $\eta_2 = \eta_3 = -1$.
In the actual crystal structure, the Fe atomic positions are slightly displaced, and the oxygen octahedra are twisted around the [111] axis, leading to a further reduction in symmetry.

\begin{figure}[tbp]
\begin{center}
\label{fig:spectra}
\includegraphics[width=8.cm]{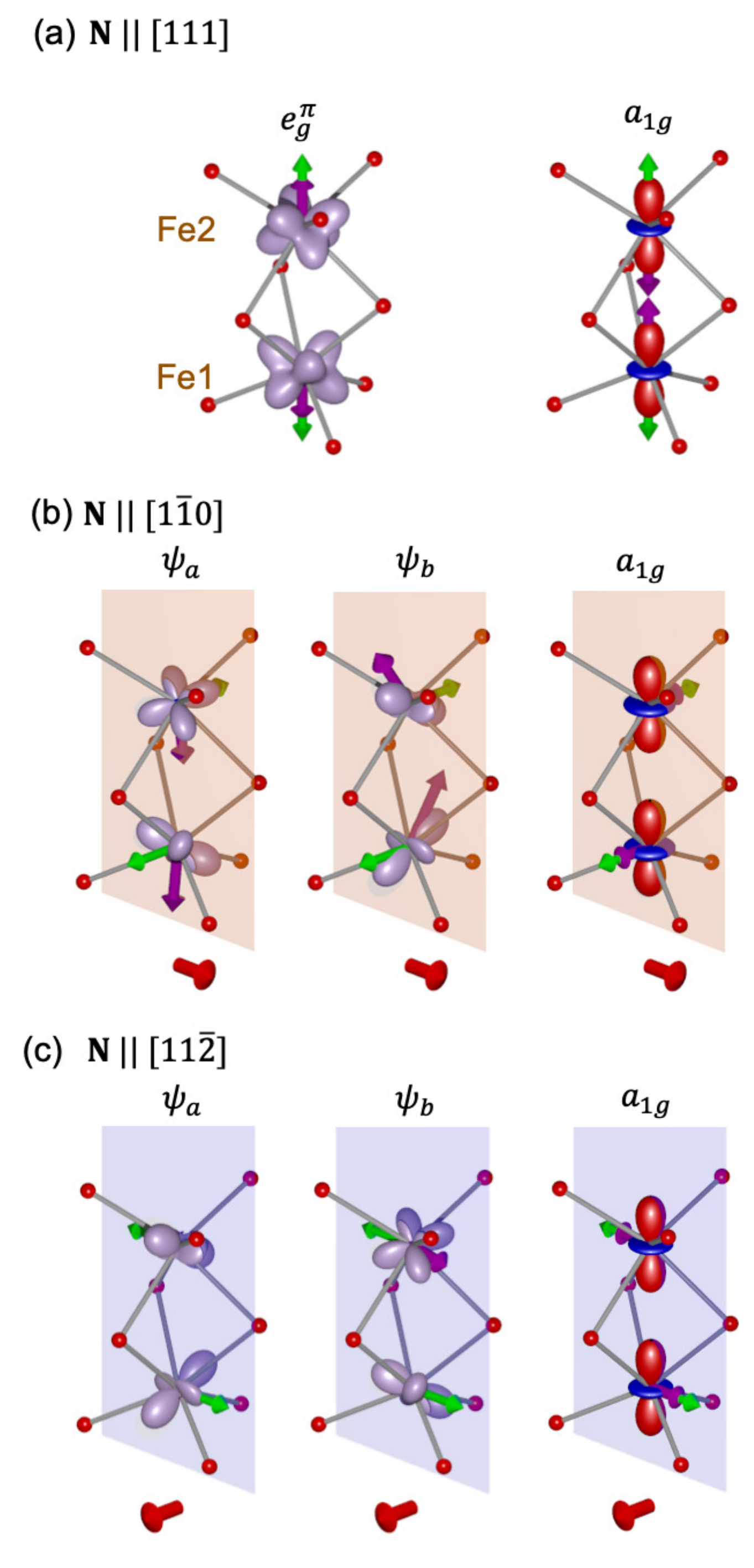}
\caption{Spin and orbital configurations 
at two lower Fe sites
for (a) $\mathbf{N}\parallel[111]$,
(b) $\mathbf{N}\parallel[1\bar{1}0]$, 
(c) $\mathbf{N}\parallel[11\bar{2}]$.
Red, blue, and purple lobes 
represent orbitals of $\psi_{a,b}$, $e_g^{\pi}$, and $a_{1g}$ electronic states.
Green, purple, and red arrows 
denote antiferromagnetic spin,  
AMD,
and ferromagnetic spin moments (${\bf M}_{\rm wFM}$), respectively.
Red and blue boards indicate $c^{\prime}$-glide and $c$-glide
planes, respectively.
} 
\end{center}
\end{figure}

Figure S1 illustrates the spatial distribution of the electronic states, calculated by the exact diagonalization of the single-ion Hamiltonian, and the expectation value of the AMD moment, $i.e.$ the $m_z^\prime$ operator described in equation below;
\begin{equation}
 \hat{\mathbf{m}}' = \frac{1}{\sqrt{10}} [3(\hat{\mathbf{r}} \cdot \hat{\boldsymbol{\sigma}})\hat{\mathbf{r}} - \hat{r}^2\hat{\boldsymbol{\sigma}}]
\end{equation}
with position operator $\hat{\mathbf{r}}$ and the Pauli matrix $\hat{\mathbf{\sigma}}$ \cite{hayami2024unified, YamasakiSTAM2025}.
The magnetic dipole term $\mathbf{T}$ in the XMCD sum rules can be expressed as $\textbf{T} = \sum_i \langle \psi_i | \hat{\textbf{m}}^\prime|\psi_i\rangle$, 
indicating that it corresponds to the sum of the local magnetic dipole expectation values for each occupied orbital $\psi_i$.
When the N\'{e}el vector $\mathbf{N}$ is parallel to the $[111]$ direction, the magnetic structure does not break the threefold rotational symmetry of the crystal; thus, the electronic states retain trigonal symmetry, preserving the $e_g^{\pi}$ and $a_{1g}$ manifolds, as shown in Fig.~S1(a).
In this situation, the local AMD becomes parallel or antiparallel to the spin moment, leading to complete cancellation.
In contrast, when $\mathbf{N}\parallel[11\bar{2}]$ or $\mathbf{N}\parallel[1\bar{1}0]$, 
the magnetic structure breaks the threefold rotational symmetry of the crystal. 
As a consequence, the $e_g^{\pi}$ manifold is further modulated by spin--orbit coupling, 
giving rise to electronic states $\psi_a$ and $\psi_b$, which produce a more anisotropic electronic distribution, as shown in Fig.~S1(b) and (c)
For $\mathbf{N}\parallel[1\bar{1}0]$, the states $\psi_a$ and $\psi_b$ exhibit finite AMD components along the [111] axis. 
The AMD components at the two Fe sites possess identical projections along [111], leading to ferroic alignment of the AMD, ie., the manifestation of an altermagnetic dipole moment.
However, for $\mathbf{N}\parallel[11\bar{2}]$, the local AMD components at the Fe1 and Fe2 sites are symmetry-opposed and cancel each other, yielding no net projection along the [111] direction.
These results establish a direct correspondence between the MPG symmetry and the electronic structure.

It should be noted that such an AMD does not emerge from an isotropic electronic state. 
For example, in the case of $t_{2g}$ orbitals, the expectation value of $t_z$ cancels out when summed over the three orbitals. 
In $\alpha$-Fe$_2$O$_3$, Fe$^{3+}$ has a high-spin $3d^5$ electronic configuration,
$|\psi_{\mathrm{gs}}\rangle = |(t_{2g}^{\uparrow})^3 (e_{g}^{\uparrow})^2\rangle$,
which is isotropic; therefore, the $T_z$ terms vanish in the ground state $\langle \psi_{\mathrm{gs}} | T_z | \psi_{\mathrm{gs}} \rangle = 0$.
However, upon excitation from the core level, the system reaches $2p^53d^6$ configurations, where the minority-spin states split into distinct energy eigenstates, $\psi_a$, $\psi_b$, and $a_{1g}$. 
For each excitation energy, the total transition probability is obtained by summing over all possible core-hole configurations, and the XMCD signal emerges from the difference between left- and right-circularly polarized absorption, which is proportional to the AMD moment \cite{YamasakiSTAM2025}. 
In practice, the spectral shape is further modulated by the Coulomb interaction between the $3d$ electrons and the core hole, resulting in a more complex lineshape.
In any case, even if the AMD vanishes in the ground state, X-ray absorption reflects the properties of the excited states; hence, XMCD originating from AMD is symmetry-allowed. 

\begin{figure}[tbp]
\begin{center}
\label{fig:spectra}
\includegraphics[width=8.cm]{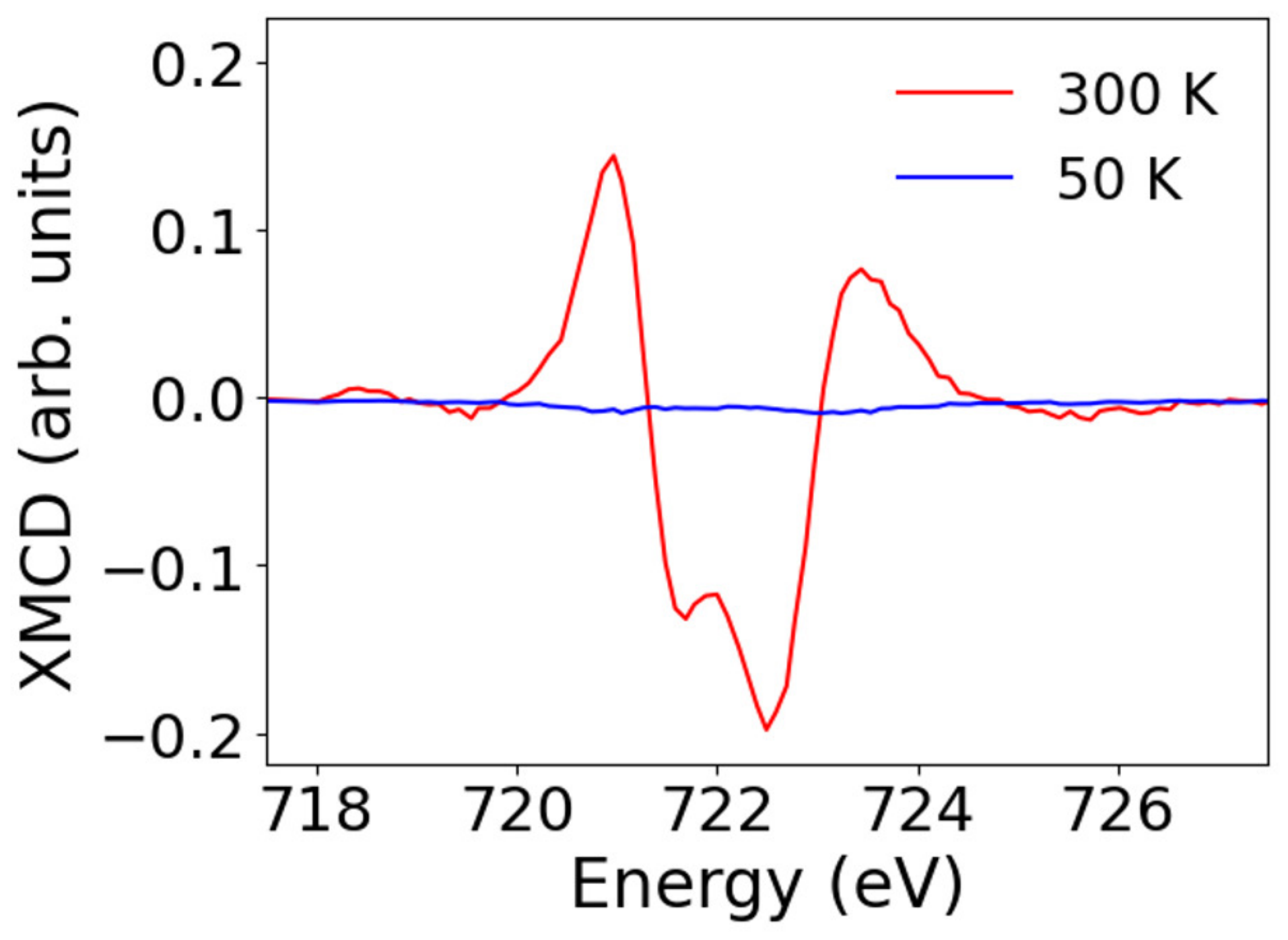}
\caption{XMCD spectra around Fe $L_2$ edge at $T = 300$ K and $T = 50$ K.
} 
\end{center}
\end{figure}

\section{XMCD spectrum below Morin Transition Temperature}
Fig.~S2 shows the XMCD spectra near the Fe $L_2$ edge measured at $T = 300$ K and $T = 50$ K. The XMCD signal vanishes at $T = 50$ K. This behavior is consistent with the fact that, below the Morin transition temperature ($T_M \approx 250$ K), $\mathbf{N}$ is parallel to the [111] direction, and the AMD moments exhibit antiferroic alignment, as illustrated in Fig.~S1(a), resulting in the disappearance of the XMCD signal.

\begin{figure}[tbp]
\begin{center}
\label{fig:spectra}
\includegraphics[width=8.cm]{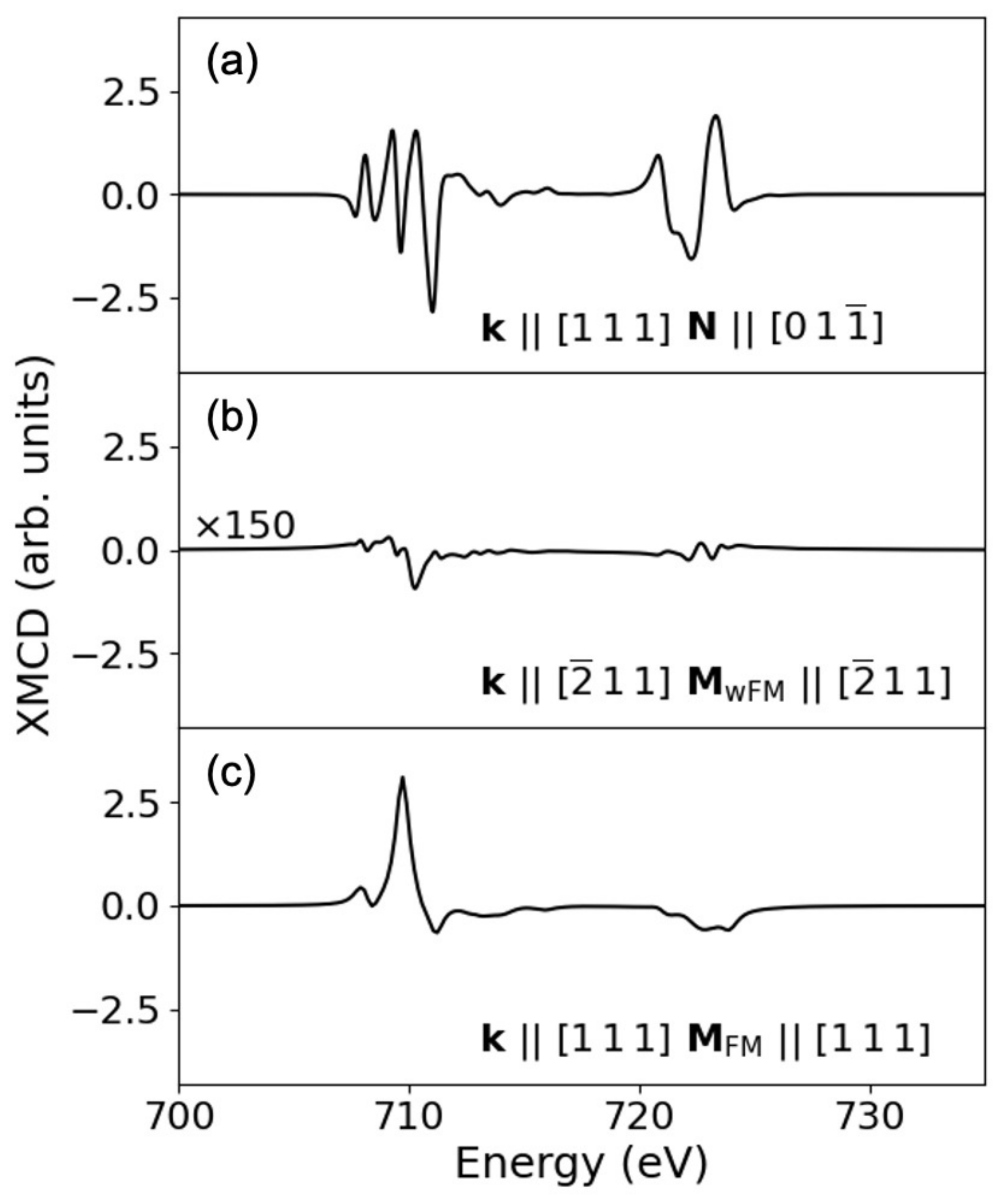}
\caption{Calculated XMCD spectra for (a) ${\bf k} \parallel [111]$ with ${\bf N} \parallel [01\bar{1}]$ and (b) ${\bf k} \parallel [\bar{2}11]$ with ${\bf M}_{\rm wFM} \parallel [\bar{2}11]$. (c) XMCD spectrum calculated for a fully ferromagnetic configuration, in which all Fe moments are aligned along the [111] direction (${\bf M}_{\rm FM} \parallel [111]$) with ${\bf k} \parallel [111]$.
} 
\end{center}
\end{figure}

\section{Comparison of XMCD spectra for antiferromagnetic, weak ferromagnetic, and ferromagnetic configurations}

\textcolor{red}{
Here, we compare the XMCD spectra associated with the antiferromagnetic, weak ferromagnetic, and ferromagnetic components. Figures~S3(a) and (b) show the calculated XMCD spectra for ${\bf k} \parallel [111]$ with $\mathbf{N} \parallel [01\bar{1}]$, and for ${\bf k} \parallel [\bar{2}11]$ with $\mathbf{M}_{\rm wFM} \parallel [\bar{2}11]$, respectively, corresponding to the XMCD spectra presented in Fig.~2(b) of the main text. 
Figure~S3(c) shows the XMCD spectrum calculated for a fully ferromagnetic configuration, in which all Fe moments are aligned along the [111] direction with ${\bf k} \parallel [111]$. The resulting spectral shape is clearly distinct from that of the XMCD spectrum shown in Fig.~S3(a). Consequently, the experimentally observed XMCD spectrum cannot be reproduced by a ferromagnetic state with the Fe moments aligned along the [111] direction.
}

\section{Multiplet calculation parameters}
\textcolor{red}{
The Fe $L_{2,3}$-edge spectra were calculated for the $2p^63d^5 \rightarrow 2p^53d^6$ transition using EDRIXS.
The ligand field was calculated from the positions of the oxygen ligands surrounding Fe, yielding a trigonal splitting of
$\Delta_{\mathrm{tri}} \equiv E(a_{1g})-E(e_g^\pi)= 7 $~meV and an effective octahedral splitting of $10Dq = 1.23$~eV, 
defined as the energy difference between the centroids of the $e_g$ and $t_{2g}$ manifolds.
The $3d$ spin--orbit coupling constants are $\zeta_{3d}=0.059$~eV in the $3d^5$ initial configuration and $\zeta_{3d}=0.074$~eV in the $2p^53d^6$ core-excited configuration.
The Fe $2p$ core spin--orbit coupling constant was set to $\zeta_{2p}=8.035~\mathrm{eV}$ in the $2p^5 3d^6$ core-excited configuration.
All $3d$--$3d$ and $2p$--$3d$ Slater integrals were scaled to 70\% of their atomic values. 
The atomic and scaled values are listed in Table~\ref{tab:slater_integrals}. 
Core-hole lifetime broadening was included using Lorentzian functions with $\Gamma$ values of 0.690~eV at the $L_3$ edge and 0.924~eV at the $L_2$ edge, corresponding to the half width of half maximum (HWHM).
The magnetic order was imposed through a local exchange-field term,
$\hat{H}_{\mathrm{ex}}^{(j)}=-\mathbf{b}_j\cdot\hat{\mathbf{S}}_j$,
where $\hat{\mathbf{S}}_j$ is the dimensionless total $3d$ spin operator at Fe site $j$ and $|\mathbf{b}_j|=0.05~\mathrm{eV}$.
The exchange-field directions were chosen to represent the magnetic configurations considered in each calculation. }

\color{red}
\begin{table}[t]
\centering
\caption{Coulomb Slater integrals used in the Fe $L_{2,3}$-edge multiplet calculation. Values are in eV.}
\label{tab:slater_integrals}
\begin{tabular}{llrr}
\hline
Configuration & Integral & Atomic value & Value used \\
\hline
$3d^5$       & $F^2(dd)$ & 12.043 & 8.430 \\
             & $F^4(dd)$ &  7.535 & 5.275 \\
$2p^53d^6$   & $F^2(dd)$ & 12.818 & 8.973 \\
             & $F^4(dd)$ &  8.023 & 5.616 \\
             & $F^2(pd)$ &  7.446 & 5.212 \\
             & $G^1(pd)$ &  5.566 & 3.896 \\
             & $G^3(pd)$ &  3.166 & 2.216 \\
\hline
\end{tabular}
\end{table}
\color{black}

\end{document}